\documentclass[9pt,a4paper,twocolumn]{article}
\usepackage[utf8]{inputenc}
\usepackage[english]{babel}
\usepackage{amsmath}
\usepackage{amsfonts}
\usepackage{amssymb}
\usepackage{graphicx}
\usepackage{fullpage}
\usepackage{comment}
\usepackage{hyperref}
\usepackage{indentfirst} 
\usepackage{authblk} 
\usepackage{comment}
\usepackage{amsfonts}
\usepackage{xcolor}
\usepackage[super,comma,sort&compress]{natbib} 

\usepackage{scalerel}
\DeclareUnicodeCharacter{2010}{-}
\DeclareUnicodeCharacter{2212}{-}

\newcommand{\rieff}{R^{eff}}
\newcommand{\fcal}{ \mathcal{F} }
\newcommand{\rvec}{ \mathbf{r} }
\newcommand{\Rvec}{ \mathbf{R} }
\newcommand{\fvec}{ \mathbf{f} }

\newcommand{\vvec}{ \mathbf{v} }
\newcommand{\reff}{R_{\text{eff}}}

\author[1]{Javier Diaz}
\author[1]{Marco Pinna\thanks{mpinna@lincoln.ac.uk}}
\author[1]{Andrei V. Zvelindovsky}
\author[2,3,4]{Ignacio Pagonabarraga\thanks{ipagonabarraga@ub.edu}}
\affil[1]{Centre for Computational Physics, University of Lincoln. Brayford Pool, Lincoln, LN6 7TS, UK}
\affil[2]{Departament de Física de la Matèria Condensada, Universitat de Barcelona, Martí i Franquès 1, 08028 Barcelona, Spain
}
\affil[3]{
{CECAM, Centre Europ\'een de Calcul Atomique et Mol\'eculaire, \'Ecole Polytechnique F\'ed\'erale de Lausanne,
Batochime - Avenue Forel 2, 1015 Lausanne, Switzerland }
}
\affil[4]{Universitat de Barcelona Institute of Complex Systems (UBICS), Universitat de Barcelona, 08028 Barcelona, Spain
}

\title{ Large scale three dimensional simulations of hybrid block copolymer/nanoparticle systems.
\footnote{Electronic supplementary information (ESI) available. }
 }

\begin{document}
\maketitle 

\begin{abstract}
Block copolymer melts self-assemble in the bulk into a variety of nanostructures, making them perfect candidates to template the position of nanoparticles. 
The morphological changes of block copolymers are studied in the presence of a considerable filling fraction of colloids.
Furthermore, colloids can be found to assemble into ordered hexagonally close-packed structures in a defined number of layers when softly confined within the phase-separated block  copolymer.
A high concentration of interface-compatible nanoparticles leads to complex block copolymer morphologies depending on the polymeric composition.
Macrophase separation between the colloids and the block copolymer can be induced if colloids are unsolvable within the matrix. 
This leads to the formation of ellipsoid-shaped polymer-rich domains elongated along the direction perpendicular to the interface between block copolymer domains.  
\end{abstract}

\section{Introduction}


Block Copolymers (BCP) are a fascinating class of materials due to their unique chain structure, made of several polymer subchains, joint covalently. 
This property translates into a rich phase behaviour in the bulk or in thin films. 
In particular, diblock copolymers have been found to self-assemble into a variety of periodic ordered structures such as lamellar, cylindrical or body-centered cubic spheres\cite{matsen_unifying_1996}. 
Colloidal particles sized a few nanoparticles are perfect candidates to  be segregated within block copolymers, which, due to their periodic structure can lead to a highly organised nanocomposite material\cite{langner_mesoscale_2012,ganesan_theory_2014}.   




Hybrid block copolymer/nanoparticle systems have been shown to co-assemble forming interesting collective behaviour that is both dependent on the polymeric properties and the number, size and chemical coating of the colloids\cite{bockstaller_size-selective_2003,bockstaller_block_2005,diaz_phase_2018,diaz_cell_2017}.
Many models have focused on the two-dimensional behaviour of such systems\cite{balazs_multi-scale_2000,pinna_modeling_2011}, which is a physically reasonable approximation as many properties can be inferred from two (2D) to three dimensions (3D). 
Moreover, many experiments and simulations are devoted to the case of thin films\cite{dessi_cell_2013,matsen_thin_1997} and ultra-thin films\cite{ploshnik_hierarchical_2013,aviv_quasi-two-dimensional_2019}. 
Nonetheless, the three-dimensional bulk assembly of block copolymers is considerably richer than the two-dimensional one. 
Similarly, the possibilities of colloidal assembly in 3D are less restricted, for instance, allowing the formation of two-dimensional layers with internal ordering.


The modification of the NP surface through, for example, grafted polymer chains, has led to a precise control of the localisation of colloids within the BCP phase-separated domain
\cite{chiu_control_2005,chiu_distribution_2007,horechyy_nanoparticle_2014} 
or the interface between them
\cite{kim_effect_2006,kim_nanoparticle_2007}. 
Since the presence of nanoparticles can induce a phase transition of the BCP, it is crucial to determine the overall morphology of the polymer nanocomposite system.

Several  computational and theoretical techniques have been used to study BCP/NP systems. 
Strong segregation theory have been used to analytically study the viscoelastic properties of polymer nanocomposites
\cite{kim_morphology_2016,pryamitsyn_strong_2006,
pryamitsyn_origins_2006}
, finding a reduction of the lamella thickness when non-selective nanoparticles are present in the interface. A lamellar to bicontinuous transition was also reported, given by the vanishing of the bending modulus of the diblock copolymer, which is in accordance with experimental findings \cite{kim_creating_2007}.  

 Monte Carlo have been used \cite{detcheverry_monte_2008,kang_hierarchical_2008} to study the assembly of BCP/NP systems on chemically nanopatterned substrates. 
 In close resemblance with experiments, this method allowed to obtain well-ordered assembled nanoparticles. 
 Additionally, Huh et al\cite{huh_thermodynamic_2000} reported the changes in the diblock copolymer morphologies due to the presence of A-compatible nanoparticles in a diblock copolymer of arbitrary morphology (that is, exploring the composition ratio) using 3D simulations. This provided a phase diagram with only a few points. 
 Molecular Dynamics \cite{schultz_computer_2005} has been used to study the phase behaviour of BCP/NP systems for different Flory–Huggins parameter values using fixed symmetric diblock copolymers.

Dissipative Particle Dynamics (DPD) \cite{liu_cooperative_2006,chen_structure_2010,
posocco_molecular_2010,
maly_self-assembly_2008}
has been used to study the aggregation of NPs within BCP melts, finding NP-assembly dependence on the lamella morphology, resulting in a transition to a complex phase. 
Self Consistent Field Theory (SCFT)
\cite{thompson_predicting_2001,thompson_block_2002,
matsen_particle_2008,lee_effect_2002,ginzburg_influence_2005}
has been widely used to study the segregation of nanoparticles within the diblock copolymer domains, reporting the size-selectivity of NP localisation found in experiments\cite{bockstaller_size-selective_2003}. 
 The Cahn-Hillard equation \cite{balazs_multi-scale_2000,ginzburg_modeling_2000,ginzburg_three-dimensional_2002} has been used to study the dynamical evolution of the phase separation, which is found to be slowed down by the presence of nanoparticles in the polymer blend. In these cases, a moderate volume fraction of nanoparticles that do not interact with each other is considered. 

Previous simulation works were using relatively modest sizes of computational boxes. 
It is known that for block copolymer systems large simulation box sizes are essential \cite{sevink_self-assembly_2005}. 
Small simulation boxes can artificially pin systems in intermediate states\cite{xu_electric_2005}.
Our previous parallel CDS scheme\cite{guo_parallel_2007} has been extended to be coupled with Brownian Dynamics for colloids, using Fortran Coarrays.  
The relative speed of this computer program allows us to reach considerably large systems, along with a high number of particles, which were previously unavailable.  
Thanks to this, we can explore a vast range of regimes, both when colloids are a mere additive and when the co-assembly of the system is driven by the dominating high concentration of colloids. 
As a result we found new phases such as, for instance, coexistence of macroscopically separated colloid rich phase and BCP lamellae, which were not observed in simulations before.

\section{Model}

The evolution of the BCP/colloids  system is determined by the excess free energy which can be separated as 
\begin{equation}
\fcal _{tot} = \fcal_{pol}+\fcal_{cc} +\fcal_{cpl}
\end{equation}
with $\fcal_{pol}$ being the free energy functional of the BCP melt, $\fcal_{cc}$ the colloid-colloid interaction and the last contribution being the coupling term between the block copolymer and the colloids. 

\subsection{Polymer Dynamics: Cell Dynamics Simulations}

The BCP is characterized by the order parameter $\psi(\rvec,t)$ which is related to the differences in the local monomer concentration $\phi_A(\rvec,t)$ and $\phi_B(\rvec,t)$ of block A and B, respectively, 
\begin{equation}
\psi(\rvec,t)=\phi_A(\rvec,t)-\phi_B(\rvec,t)+(1-2f_0)
\end{equation}
with the composition ratio $f_0=N_A/(N_A+N_B)$ being the overall volume fraction of monomers A in the system. $\psi(\rvec,t)$ is considered the local order parameter, which has a value $0$ for the disordered-or homogeneous- state and $|\psi|>0$ for microphase-separated regions. 

The time evolution of $\psi(\rvec,t)$ is dictated by the conservation of mass, resulting in the Cahn-Hilliard-Cook equation \cite{cahn_free_1959,cook_brownian_1970}
\begin{equation}
\frac{\partial\psi ( \rvec, t )}{\partial t}=
M \nabla^2 \left[
\frac{\delta F_{tot} [ \psi] }{ \delta \psi}
\right]+
\eta ( \rvec, t)
\label{eq:cahn}
\end{equation}
with $M$ being a mobility parameter and $\eta(\rvec,t)$ being a gaussian noise parameter that satisfies the fluctuation-dissipation theorem
\begin{equation}
\langle \eta(\rvec,t) \eta(\rvec',t')\rangle =
-k_B T M \nabla^2 \delta(\rvec-\rvec')
\delta(t-t')
\end{equation}
for which we have used the algorithm given by Ball\cite{ball_spinodal_1990}. $k_BT$ sets the thermal energy scale of the diblock copolymer. 

The total free energy present in Equation \ref{eq:cahn} is decomposed into purely polymeric, coupling and intercolloidal free energy, respectively, 
\begin{equation}
F_{tot}=
F_{OK}+F_{cpl}+F_{cc}
\end{equation}
where the purely polymeric free energy $F_{OK}$ is the standard Ohta-Kawasaki free energy \cite{ohta_equilibrium_1986}. Furthermore, the diblock copolymer free energy $F_{OK}=F_{sr}+F_{lr}$ can be decomposed in short ranged
\begin{equation}
\label{eq:Fshort}
F_{\text{sr}}[\psi]=\int d\rvec 
\left[ 
H(\psi)+\frac{1}{2} D |\nabla\psi|^2 
\right]
\end{equation}
and long-ranged free energy, 
\begin{equation}
\label{eq:Flong}
F_{lr}[\psi]=
\frac{1}{2} B\int d\rvec \int d\rvec'
G(\rvec,\rvec')\psi(\rvec)\psi(\rvec')
\end{equation}
with $G(\rvec,\rvec')$ satisfying $\nabla^2 G(\rvec,\rvec')=-\delta(\rvec-\rvec')$,i.e., the Green function for the Laplacian. 

The local free energy can be written as  \cite{hamley_cell_2000}
\begin{equation}
H(\psi)=
\frac{1}{2}\tau'\psi^2 
+\frac{1}{3} v(1-2f_0)\psi^3 +\frac{1}{4} u \psi^4
\end{equation}
where $\tau'=-\tau_0+A(1-2f_0)^2$, $u$ and $v$ can be related to the molecular structure of the diblock copolymer chain \cite{ohta_equilibrium_1986}. 
The local free energy $H(\psi)$ possesses 2 minima values $\psi_{-}$ and $\psi_{+}$ which are the values that $\psi(\rvec,t) $ takes in the phase-separated domains.  
Parameter $D$ in Equation \ref{eq:Fshort} is related to the interface size $\xi=\sqrt{D/\tau'}$ between domains and $B$ in Equation \ref{eq:Flong} to the periodicity of the system $H\propto 1/\sqrt{B}$ as the long ranged free energy takes into account the junction of the two chains in a diblock copolymer.

Contrary to the block copolymer -which is described continuously- NPs are individually resolved. 
We consider a suspension of $N_p$ circular colloids with a tagged field moving along its center of mass $\psi_c(r)$.
 The presence of  nanoparticles in the BCP is introduced by a coupling term in the free energy,  which takes a simple functional form
\begin{equation}
F_{cpl}[\psi,\{\Rvec_i\}]= 
\sum_{p=1}^{N_p}
\sigma\int d\rvec\ \psi_{c}\left(\rvec-\Rvec_p \right)
\left[\psi(\rvec,t)-\psi_0    \right]^2
\end{equation}
with $\sigma$ a parameter that controls the strength of the interaction and $\psi_0$ an affinity parameter that is related to the preference of the NP towards different values of the order parameter $\psi(\rvec,t)$.
 A particle with an affinity $\psi_0=1$ is purely coated with copolymer A while a mixed brush would result in $\psi_0=0$. 
$\psi_c(\rvec)$ is a tagged function that accounts for the size and shape of the nanoparticle. 
At the same time, it can be tuned to define a soft and a hard-core for the nanoparticle regarding the coupling with the BCP. 
In our simulations we use 
\begin{equation}
\psi_{c} (\rvec) = 
\exp\left[
1-\frac{1}{1-\left( \frac{| \rvec   |}{\rieff}  \right)^\alpha} 
\right]
\label{eq:psici}
\end{equation}
from which we obtain a relationship $\reff = R_0  \left( 1+1/\ln 2    \right)^{1/\alpha} $ such that the tagged field has been reduced to $0.5$ at $r=R_0$. 
$R_{eff}$ also acts as the cut-off distance in the coupling interaction, i.e., $\psi_c(r>R_{eff})=0$.
We select $\alpha =2$ to provide a smooth decay in $\psi_c(r)$.

Nanoparticles are considered soft in their interparticle interaction, following a Yukawa-like potential 
\begin{equation}
U(r)=
U_0 
\left[\frac{ 
\exp\left(1-r/R_{12}   \right)}{r/R_{12}}-1
\right]
\end{equation}
with $R_{12}=2R_0$ and $r$ being the center-to-center distance.  

 Colloids undergo diffusive dynamics, described by the Langevin equation in the over damped regime. The center of mass of each colloid $\Rvec_i$ is considered to follow Brownian Dynamics, that is,
\begin{equation}
\label{eq:brownian}
\vvec_i=
\frac{1}{\gamma} \left(
\fvec^{c-c}+\fvec^{cpl}+\sqrt{2k_BT\gamma}\xi
 \right)
\end{equation}
with $\gamma$  the friction coefficient, $k_BT$ is the NP thermal energy and $\xi$ is a random gaussian term satisfying fluctuation dissipation theorem. The coupling force $\fvec_i^{cpl}=-\nabla F_{cpl}$ accounts for the interaction between the nanoparticle and the BCP medium. 
Similarly, $\fvec_i^{cc}=-\nabla F_{cc}$.

The order parameter time evolution presented in Equation \ref{eq:cahn} is numerically solved using a cell dynamic simulation scheme\cite{oono_study_1988,bahiana_cell_1990}, for which the laplacian is approximated as $\frac{1}{\delta x ^2}  [ \langle\langle X \rangle\rangle -X  ] $ with 
\begin{equation}
\langle\langle \psi \rangle\rangle  = 
\frac{6}{80}  \sum_{NN}  \psi   +
\frac{3}{80} \sum _{NNN} \psi+
\frac{1}{80} \sum _{NNNN} \psi
\end{equation}
in three-dimensional systems. NN, NNN and NNNN stand for nearest-neighbour, next-nearest-neighbour and next-next-nearest neighbour,  respectively, that is, summation over lattice points around the lattice point $\psi_{ij}$. 
The lattice is characterized by its spacing $\delta x $. 

We have extended the previous CDS parallel scheme\cite{guo_parallel_2007} into hybrid CDS/Brownian Dynamics parallel implementation using Fortran Coarrays\cite{fanfarillo_opencoarrays:_2014}. 
This allow us to study large system sizes and long time behaviour which is essential to understand the ordering and mesoscopic properties that will be studied in this work.  

In order to study the ordering of colloids and distinguish between fluid and solid-like structures, we make use of the three-dimensional bond order parameter $Q_n$\cite{rein_ten_wolde_numerical_1996}, which takes finite positive values for the case of cubic -$Q_4$- or hexagonal close packed (HCP) -$Q_6$- configurations and both vanishing in the fluid regime.

\section{Results}

We aim to study the three dimensional phase behaviour of block copolymers in the presence of colloidal nanoparticles along with the assembly of colloids. 
Following a complete two-dimensional description of the main parameters in play \cite{diaz_phase_2018} we consider representative 3D cases of special interest. 
In this work we use the standard parameters of CDS: $\tau_0=0.35$, $A=1.5$, $u=0.5$, $v=1.5$ $M=1$  and $D=1$.
 The size of the block copolymer period is controlled by $B=$ which takes a value of $B=0.002$ unless stated otherwise.
 The thermal scale is set to $k_B T = 0.1$ and the friction constant follows $\gamma  = 6\pi R_0 \eta_0$ with viscosity given by $\eta_0 = 0.1$. 
 The coupling strength and the NP-NP interaction scale are set to $\sigma=1$ and $U_0=1$, respectively.  
The time and length discretization are set to $\delta t =0.1$  and $\delta x =1.0$.   
Time scales will be expressed in units of the diffusion time of the BCP $\tau_{pol} = \xi^2/(\tau_0 M)$ with $\xi$ being the micro-phase separation length scale. 
Unless otherwise specified, simulations were performed in $64^3$ box sizes.

\subsection{NPs compatible with one copolymer }

\subsubsection{Phase transition induced by NPs}

One of the most common instances of NP dispersion in BCP is the case of colloidal particles which are coated to be compatible with one of the blocks, for example, when coated with the same homopolymer A in a A-\textit{b}-B diblock copolymer. 
Such particles have been found to segregate to their preferred domain, both experimentally\cite{ploshnik_hierarchical_2010} and in simulations\cite{ginzburg_three-dimensional_2002}. 
Nonetheless, many works have shown that a considerable concentration of such particles can induce phase transition due to the swelling of hosting domains (in this case, A phase). 
A full phase diagram has been achieved using Monte Carlo\cite{huh_thermodynamic_2000}  simulations and  CDS in two dimensions \cite{diaz_phase_2018}. 
In this section we will explore both the morphological transition as the  number of particles is increased, as well as the time evolution starting from a disordered block copolymer.

\begin{figure*}[h]
\centering
\includegraphics[width=0.9999\textwidth]{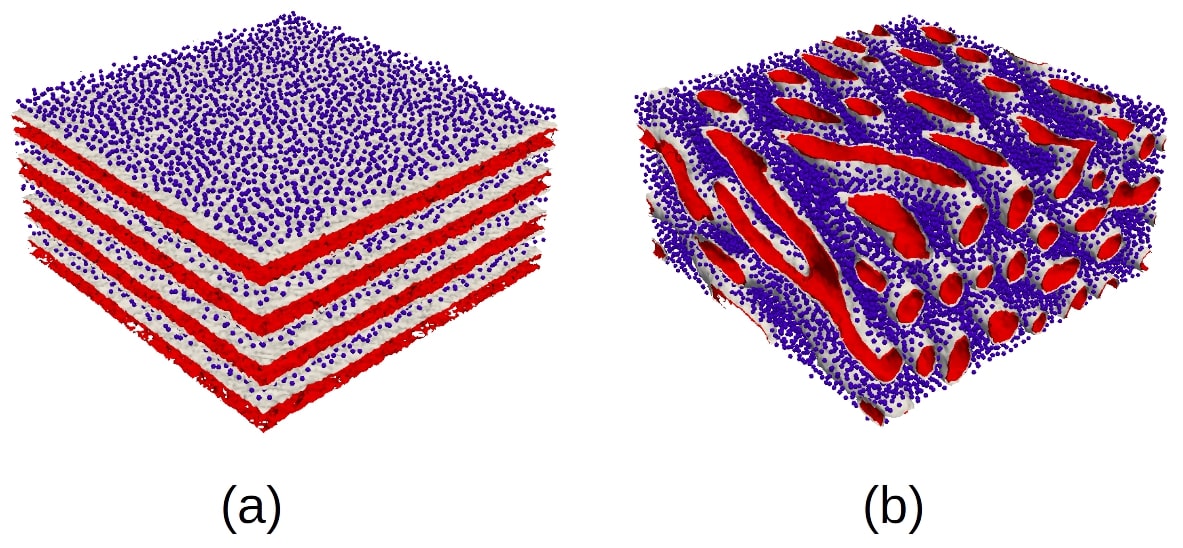}
\caption{Transition in a symmetric diblock copolymer mixture ($f_0=1/2$) induced by the presence of a volume fraction of colloids $\phi_p$. Nanoparticles are compatible with the grey phase. 
The block copolymer melt is initialised as a sinusoidal in the Z direction. 
Concentrations are $\phi_p=0.15$ and $\phi_p=0.47$ for (a) and (b), respectively. 
Simulation box is $128\times128\times 64$ grid points. 
Colloids are shown as spheres in blue. 
The block copolymer is visualised through isosurfaces of $\psi(\rvec)$ field, showing the A (red) and B (gray) domains.  
  }
\label{fig:3d.transition_iso2c}
\end{figure*}

 NPs with a radius $R_0=1.5$ and symmetric block copolymer $f_0=0.5$ are initially ordered in a lamellar morphology
Then, the system with a number of particles $N_p$ is let to evolve for a time $t/\tau_{pol} = 48.3 \times 10^3$ in terms of the polymer diffusion time. 
Although this final configuration cannot be assured to be the equilibrium structure, it is a representative steady state of the system, after visual inspection and analysis of parameters such as $<|\psi(\rvec,t)|>$, as described by Ren et al\cite{ren_cell_2001}. 
At low concentrations of particles $\phi_p=0.15$,  colloids are simply located within their preferred domain, which in this case is a simple horizontal lamellar domain, as in Figure \ref{fig:3d.transition_iso2c} (a). 
A larger number of particles results in a stronger confinement of colloids within their preferred phase, which eventually leads to a break-up of the lamellar structure into a cylindrical phase, as can be seen in Figure \ref{fig:3d.transition_iso2c} (right) for a relatively high concentration of $\phi_p=0.465$.

\begin{figure*}[h!]
\centering
\includegraphics[width=0.8\textwidth]{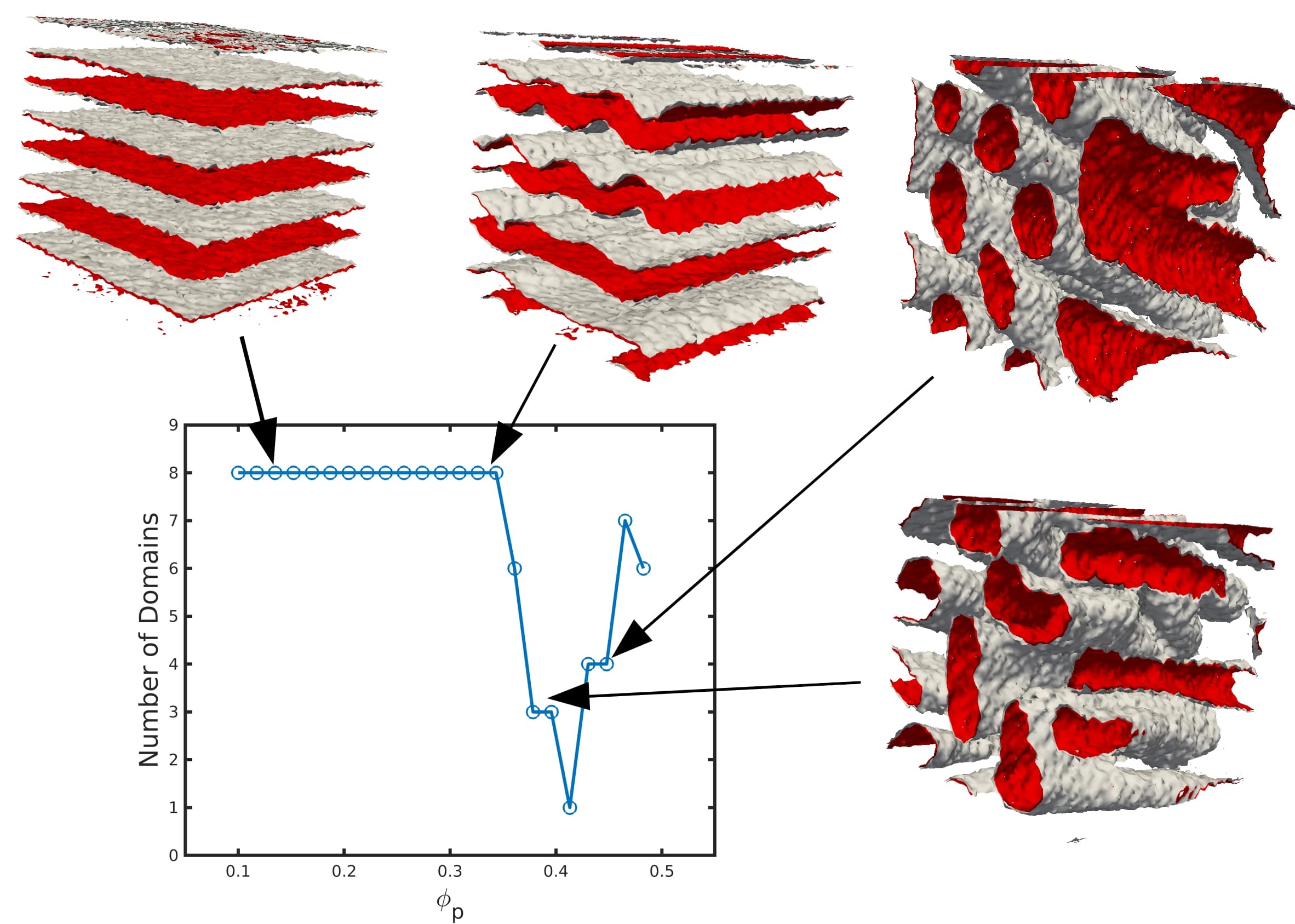}
\caption{Number of domains of a symmetric ($f_0=1/2$) block copolymer as a function of the concentration of colloids $\phi_p$. 
Snapshots of the final state of some representative simulations are shown, where the colloids are not shown for clarity. 
Simulation box is $V=64^3$
   }
\label{fig:3d.Ndomains}
\end{figure*}

We can gain insight over both the dynamical evolution of such a transition, as well as the equilibrium transition for several values of the number of particles. 
Firstly, Figure \ref{fig:3d.Ndomains} shows the number of block copolymer domains for different values of the colloidal concentration $\phi_p$.
The BCP domains can be calculated by identifying the interface between A and B domains.  
The BCP morphology is consistently lamellar characterized by $8$ domains (result of $4$ BCP periods). 
As the gray domains are filled with particles the width of the domains is enhanced, up to a transition concentration  $\phi_p^*\sim 0.35$. 
In the transition point the lamellar interface is no longer flat and presents undulation in order to further accommodate the concentration of NPs.  
After that, the number of BCP domains abruptly drops, which hints of a bi-continuous morphology of well-connected domains. 
 As the concentration of particles is again increased we observe how the block copolymer transitions again into cylindrically-shaped domains with hexagonal packing. 
The nanoparticles are effectively increasing the fraction of A monomers in the system, which is  equivalent to exploring a horizontal deviation in the $f_0-\chi N$ phase diagram for pure block copolymer melts\cite{matsen_unifying_1996}.

\begin{figure*}[h!]
\centering
\includegraphics[width=0.999\textwidth]{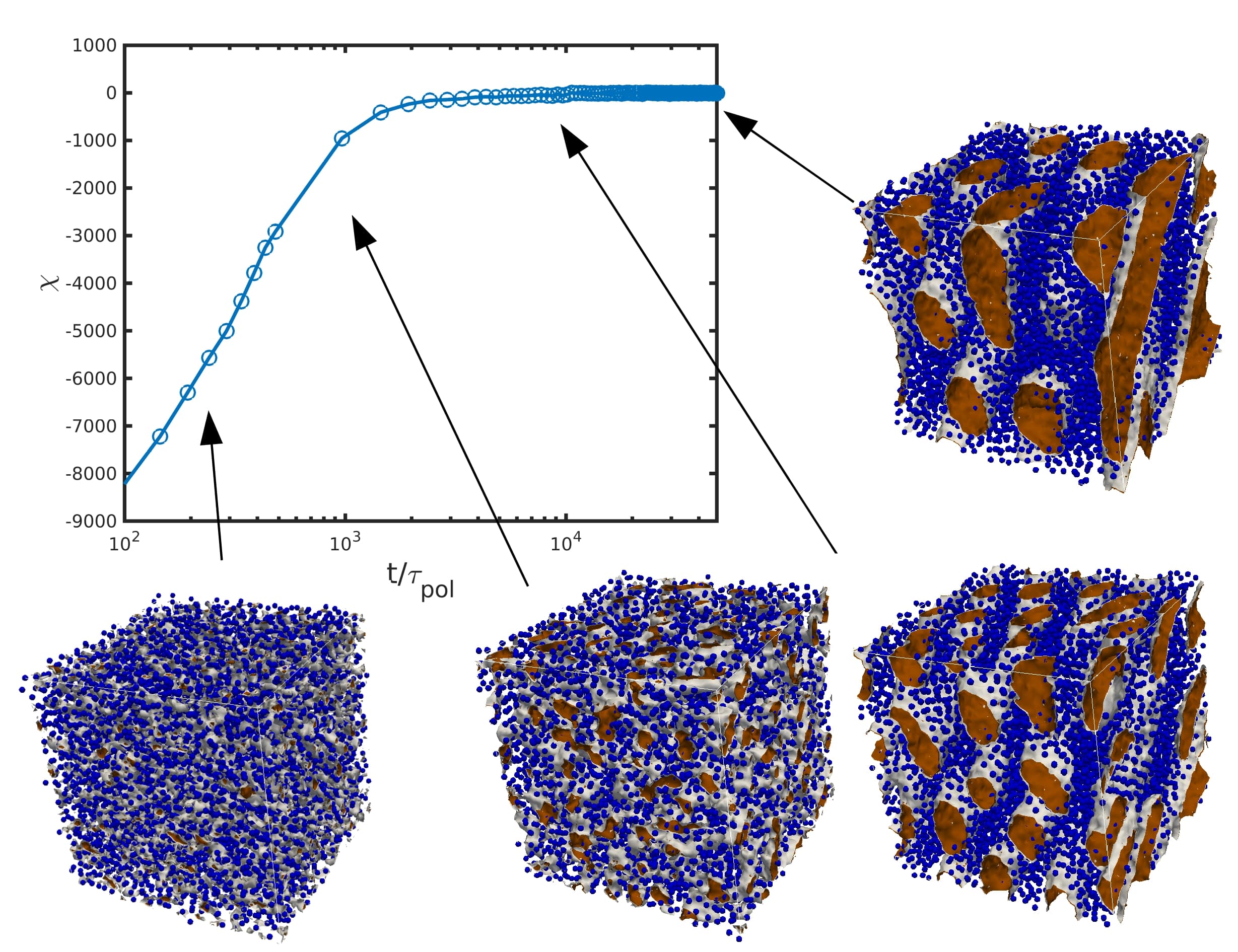}
\caption{Time evolution of an initially-disordered symmetric BCP ($f_0=1/2$) mixed with a concentration $\phi_p=0.465$ of colloids, tracked by the Euler characteristic $\chi$. Snapshots show specific simulation steps.  }
\label{fig:3d.longtime.evolution}
\end{figure*}

Secondly, we can track the kinetic pathway to equilibrium following a quench from a disordered state, both for the block copolymer and the NPs. 
Figure \ref{fig:3d.longtime.evolution} shows the Euler characteristic of the block copolymer in time, for the same parameters as Figure \ref{fig:3d.Ndomains} and $\phi_p=0.465$. 
The time evolution of $\chi$ is in fact equivalent to the reported one for a cylinder-forming pure BCP melt\cite{sevink_kinetic_2004}, thus confirming the role of nanoparticles as effectively increasing the hosting monomer fraction. 
The Euler characteristic   can be seen to approach $\chi \to 0$ from negative values which indicates the formation of isolated cylinders from a connected network of cylindrical domains.
Bottom-right and right snapshots confirm this assertion. 

\subsubsection{Assembly of colloids }

\begin{figure*}[hbtp]
\centering
\includegraphics[width=0.99\textwidth]{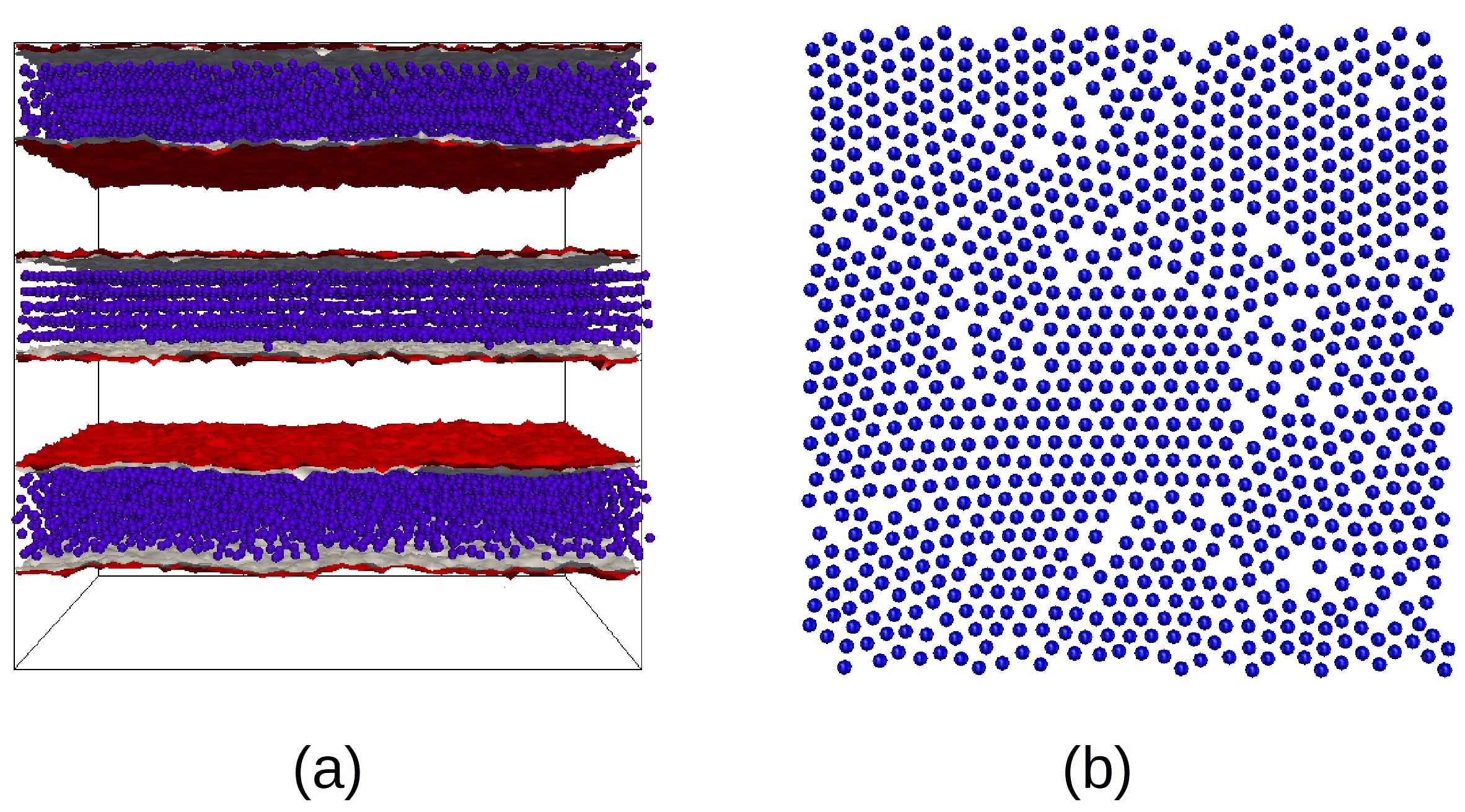}
\caption{Example of colloidal ordering assembly in a soft confinement in the presence of block copolymer. In (a) the interface between A-B BCP is shown and a frontal view of the colloids is presented. 
In (b) we can observe a top view of a single layer of the hexagonal organisation of colloids where the block copolymer is not depicted for clarity. }
\label{fig:3d.assembly.example}
\end{figure*}

Conversely,the assembly of colloids under block copolymer confinement can be studied. 
A value of $f_0=0.39$ is chosen, such that phase transition from lamella to cylinders is prevented\cite{diaz_phase_2018}. 
An initially phase-separated block copolymer and initially disordered colloidal set of $N_p$ particles are chosen with a radius $R_0$. 
As an example, Figure \ref{fig:3d.assembly.example} shows the soft confinement induced by three lamellar domains in the colloids. 
A layered organisation can be hinted in (a) while the hexagonal packing of the layers colloids within each layer can be observed in (b).

\begin{figure}[h]
\centering
\includegraphics[width=0.999\linewidth]{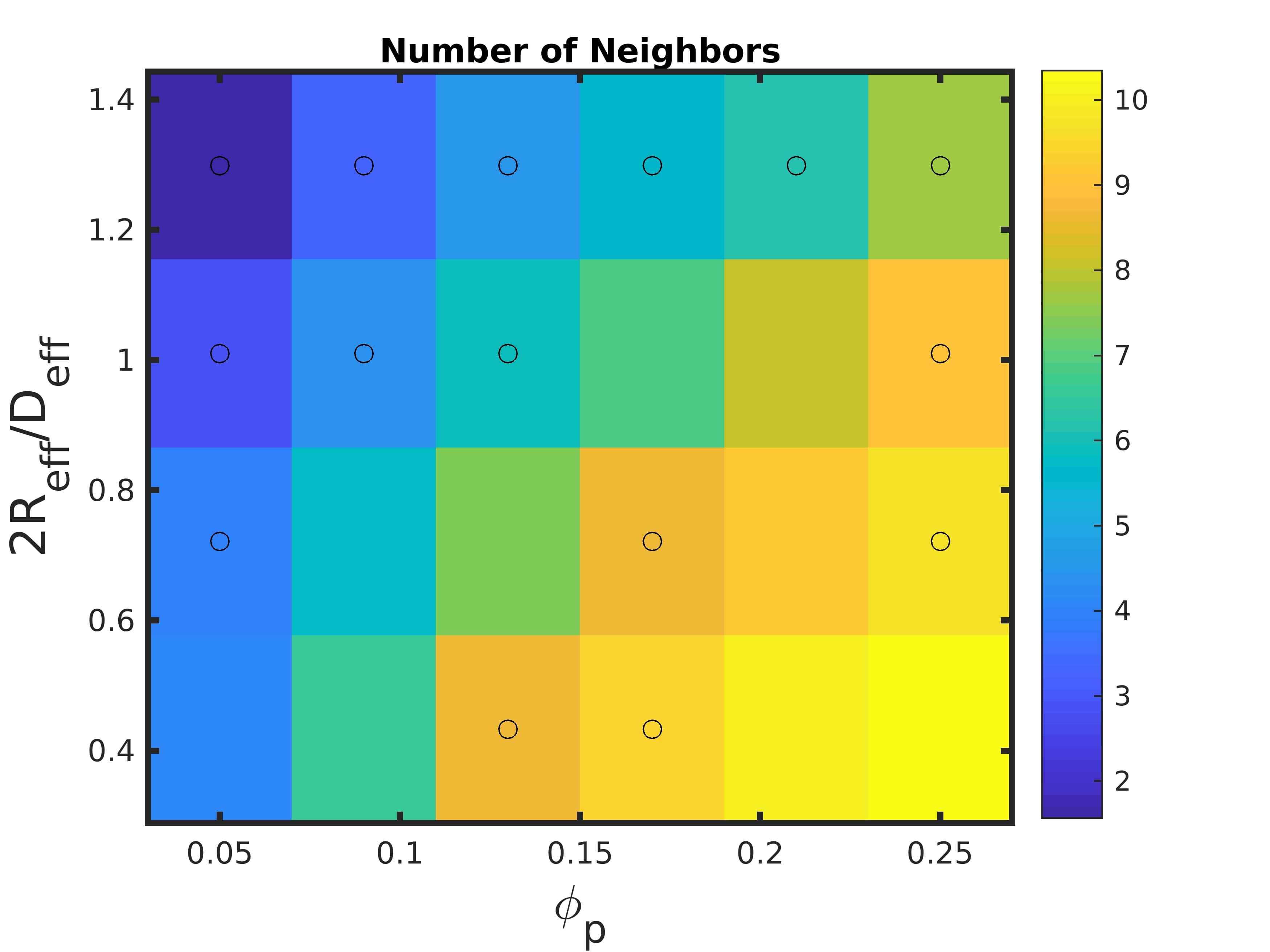}
\caption{Color map of the average number of colloidal neighbors for each simulation of an asymmetric block copolymer mixed with minority-compatible nanoparticles.
 Circles represent simulation points such that $Q_6>0.15$. }
\label{fig:3d.Nneigh}
\end{figure}

A-compatible colloids in a lamellar-forming block copolymer are softly confined, as opposed to a hard confinement between parallel plates. 
In order to have an a priori estimation of the confinement effect, in Figure \ref{fig:3d.Nneigh} we plot the reduced diameter of the particle $2R_{eff} / D_{eff}$ where $R_{eff}$ is simply the soft-core radius of the particle(eg. the cut-off of the BCP-NP interaction), while $D_{eff}=H/2-2\xi$ serves as an estimate of the spacing available for the particle, where $H/2$ is half a lamellar period while $2\xi$ is the thickness of the interface.   

In Figure \ref{fig:3d.Nneigh} we can find the average number of first colloidal neighbours  for each simulation in a color map  (see colorbar on the right).
Both the concentration of NPs and the particles size are explored.   
Visual inspection of the simulation results confirms the assembly of colloids in different number of layers, growing with the concentration of particles present in the system. 
One can then conclude that an increasing number of particles forces a close-packed type of assembly with increasingly larger number of colloidal layers. 
Moreover, the relative size of particles and lamellar spacing dictates the rate of growth in the number of layers.

Detailed insight over the ordering of colloids within these layers can be obtained by using the 3D hexagonal close-packing (HCP) order parameter $Q_6$ which is $0.75$ for a perfect HCP configuration\cite{rein_ten_wolde_numerical_1996}. 
Similarly, $Q_4$ characterizes the cubic structure. 
In Figure \ref{fig:3d.Nneigh} points for which $Q_6>0.15$ are shown, that is, systems in which the ordering is above that of a disordered liquid. 
At first sight, one could suspect that the behaviour of $Q_6(\phi_p)$ is non-monotonic, as for a fixed colloidal size it reaches higher values to then decreases.

 \begin{figure}[hbtp]
 \centering
 \includegraphics[width=0.999\linewidth]{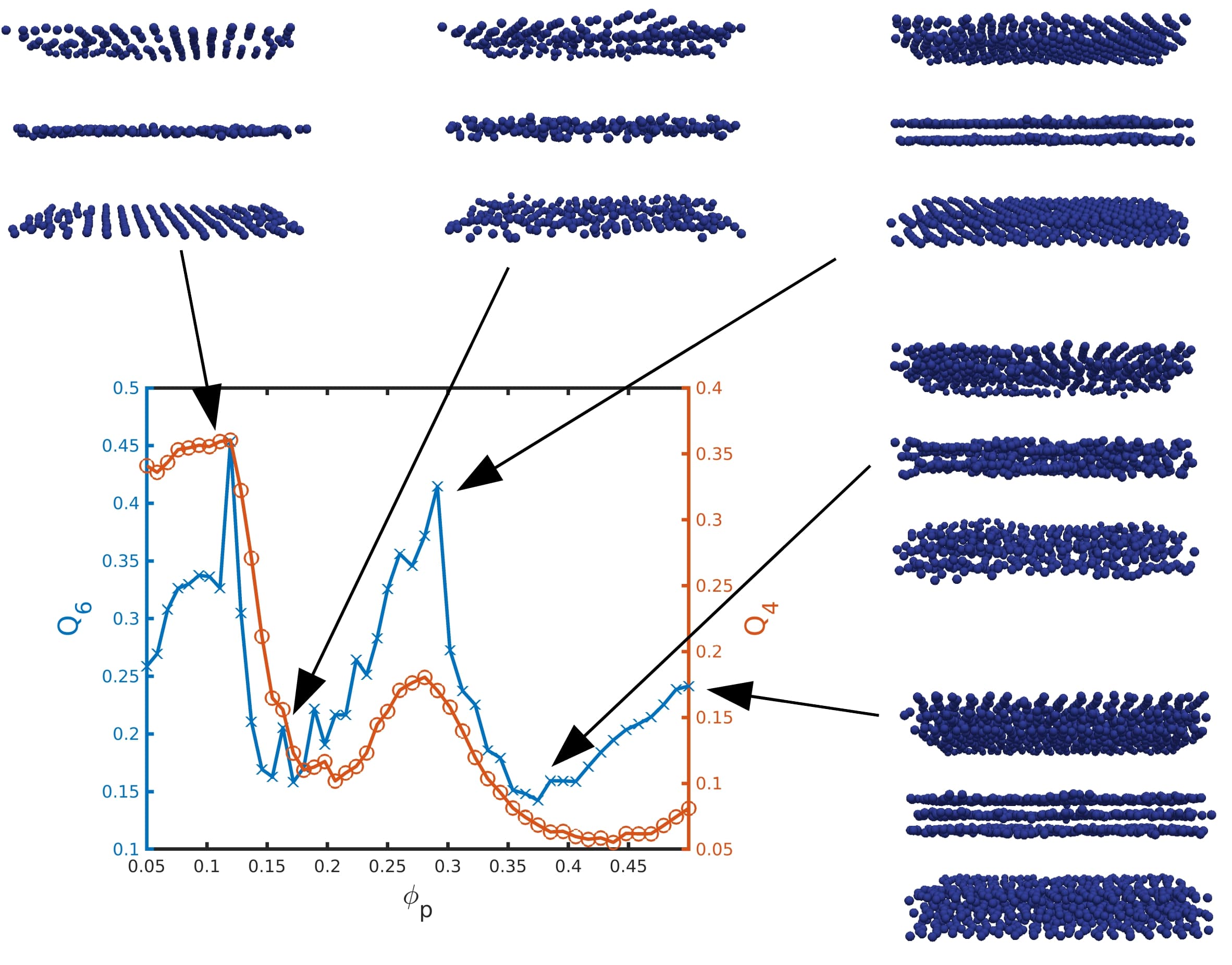}
 \caption{$Q_6$ (left axis, blue $\times$)  and $Q_4$ (right axis, red $\circ$) order parameters of the colloidal assembly in terms of the fraction of particles in the system $\phi_p$. Snapshots of the final configurations are shown with arrows indicating its respective simulation points in the $Q_6$ curve. 
 Images of the block copolymer are missing in order to help the visualisation of the colloidal horizontal, in-domain, ordering.  }
 \label{fig:3d.Qn-phip}
 \end{figure}
 
In order to study the behaviour of $Q_n$ and the assembly of colloids, we can focus on an specific particle size $R_0=2.33$   and calculate $Q_n$ for a range of concentrations. 
In Figure \ref{fig:3d.Qn-phip} we can see the  non-monotonic behaviour of both order parameters. 
At low concentration  $Q_6$ approximately grows with $\phi_p$ while $Q_4$ remains constant. 
A considerable positive value of $Q_6$ indicates a degree of interparticle ordering, which is due to the particle size being large enough to induce an effective particle attraction. 
This is rather weak at low concentrations, but as the 2D monolayers are filled, a close-packing entropic interaction results in a broad peak in $Q_6$. 
The addition of higher number of particles, instead, does not produce increased ordering but destroys the monolayer structure. 
This results in a sharp decrease in $Q_6$ and $Q_4$ that is followed by a steady increase in $Q_6$, since a similar behaviour is occurring as in the case of a monolayer, only now we have close-packed hexagonal ordering in two layers of colloids. 
 A sharp decrease is again followed by the formation of a three-layer with considerable colloidal ordering.

\begin{figure}[hbtp]
 \centering
 \includegraphics[width=0.999\linewidth]{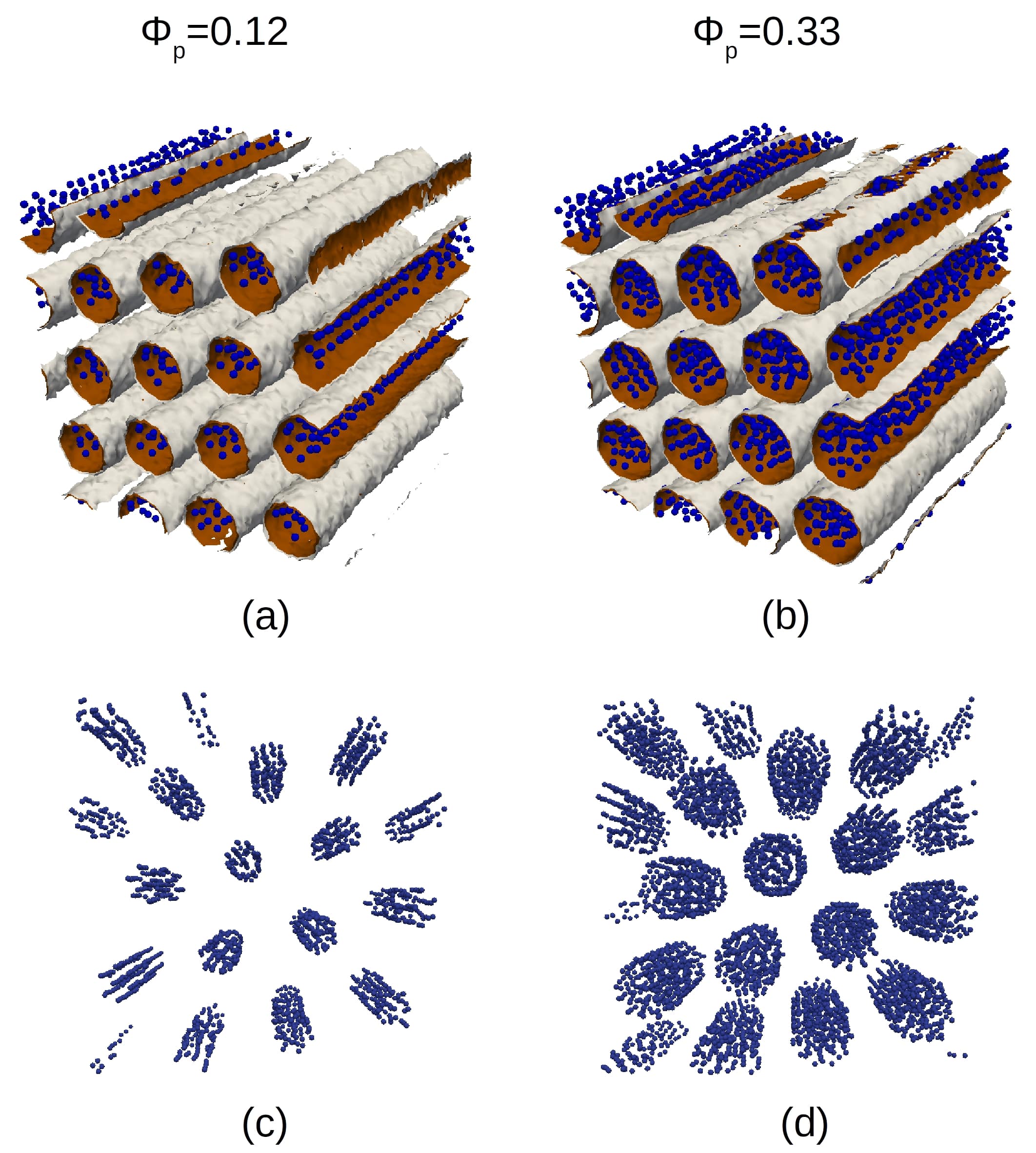}
 \caption{Mixture of cylinder-forming BCP ($f_0=0.35$)  and minority compatible nanoparticles at two different concentrations (a) and (b). 
 (c) and (d) correspond to the frontal view of (a) and (b), respectively, showing only the colloidal assembly. 
 }
 \label{fig:3d.cyl.bulk.snaps}
 \end{figure}

 
 A similar behaviour can be found in a mixture of asymmetric($f_0=0.35$), cylinder-forming block copolymer and minority-compatible nanoparticles($\psi_0=-1$). 
 Figure \ref{fig:3d.cyl.bulk.snaps} shows the BCP/NP co-assembly in cylinders at low concentration ($\phi_p=0.12$, (a)) and large concentration ($\phi_p=0.33$, (b)). 
While the cylindrical morphology is preserved in the BCP, the nanoparticles enhance the size of the cylinders without changing the number of domains. 
In fact, the nanoparticles are confined within the red, minority phase forming arrays of particles along the direction of the cylinders. 
Figure \ref{fig:3d.cyl.bulk.snaps} (c) and (d) show the frontal view of such an arrangement corresponding to the top (a) and (b) snapshots. 
A radial assembly of colloids along the axis of the cylinders is equivalent to the layered configuration in Figure \ref{fig:3d.assembly.example}, in which we again observe an increased number of layers with higher concentration. 
 
\subsection{Interface-compatible NPs}

Nanoparticles which are grafted with a mixed brush of homopolymer can be made compatible with the interface between block copolymer domains (neutral NPs)\cite{kim_creating_2007}. 
Experimentally, neutral NPs have been found to segregate to the interface between block copolymer domains. 
At high concentrations, a symmetric block copolymer can undergo lamellar-to-bicontinuous phase transition\cite{kim_nanoparticle_2007,kim_tailoring_2009}. 

We can study an initially-disordered BCP melt with a volume fraction $\phi_p$ of particles with a neutral affinity $\psi_0=0$. 
The nanoparticle size $R_0=1.5$ as well as the BCP length scales -$D=0.503$ and $B=0.02$- are chosen, following a complete phase diagram exploration of the length scales of the system, which can be found in Figure S1  in the Supplementary Information. 
In Fig. \ref{fig:3d.lam-neutral} (a) the aggregation of NPs in a lamellar-forming BCP can be observed.
The concentration of NPs is relatively low $\phi_p=0.1$.  
This aggregation is driven by the minimisation of the total free energy by creating a NP-rich domain where the order parameter takes an almost zero $\psi\approx 0$ value. 
This is, in fact, a disordered area induced by the coating of the NP.

In Fig. \ref{fig:3d.lam-neutral} (b) a $\phi_p=0.24$ concentration of NPs is shown. 
For this  higher concentration the colloids are able to build an almost continuous network of NP-rich domains. 
Macrophase separation is occurring, which can be confirmed in  Fig. \ref{fig:3d.lam-neutral.details} (a), where we run a simulation in a small box $V=32^3$ and $\phi_p=0.27$ with two distinct domains are formed. 
Additionally, the orientation of lamellar domains with respect to the NP-rich domain is clearly normal with a small tilted effect which is  due to the box size effect. 
The normal orientation is characteristic of a neutral wall\cite{pinna_block_2010}.

In Fig. \ref{fig:3d.lam-neutral} (c) the concentration of NPs is high enough -$\phi_p=0.45$- to create a single NP-rich matrix in which the BCP assembles into separated domains. 
A detail of the simulation can be found in Fig. \ref{fig:3d.lam-neutral.details} (b), where the NPs are not shown for clarity. 
In order to favour a perpendicular to contact angle between the lamellar planes and the BCP/NP boundary, the BCP-rich domains are ellipsoidal shaped. 
This has been shown in Ref. \cite{pinna_block_2008} in the context of nanoshells. 
For these reasons, the BCP-rich domains minimises the free energy by maximising the contact surface between the alternating domains and the NP-rich matrix, and minimising the exposure of A-rich or B-rich only domains to the matrix. 
This results into isolated, elongated domains which are growing in the direction normal to the interface between domains.  
In Figure S2 (bottom-left) of the Supplementary Information we can observe the 2D counterpart of ellipsoid-shaped BCP-rich domains. 
These can be again directly compared to block copolymer/homopolymer blends as in Figure 4 in Ref \cite{ohta_dynamics_1995}. 

\begin{figure}[hbtp]
\centering
\includegraphics[width=0.9999\linewidth]{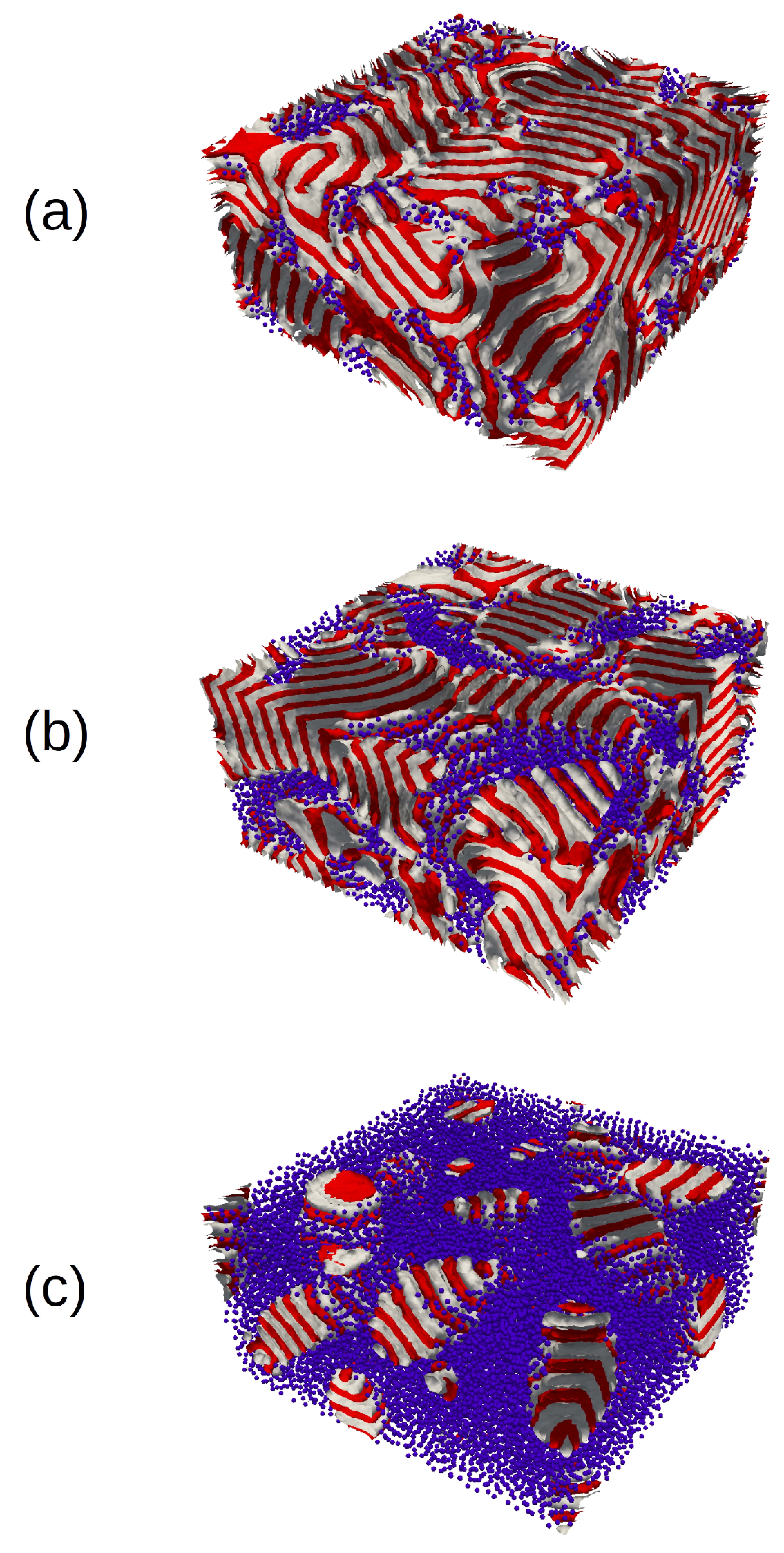}
\caption{Phase transition of a symmetric ($f_0=0.5$) BCP induced by the presence of a concentration $\phi_p$ of neutral nanoparticles. 
The concentrations of NPs are $\phi_p=0.1$, $0.24$ and $0.45$ for (a), (b) and (c), respectively.  
System size is $V=128^2\times 64$ using $8$ processors. 
}
\label{fig:3d.lam-neutral}
\end{figure}

\begin{figure}[hbtp]
\centering
\includegraphics[width=0.999\linewidth]{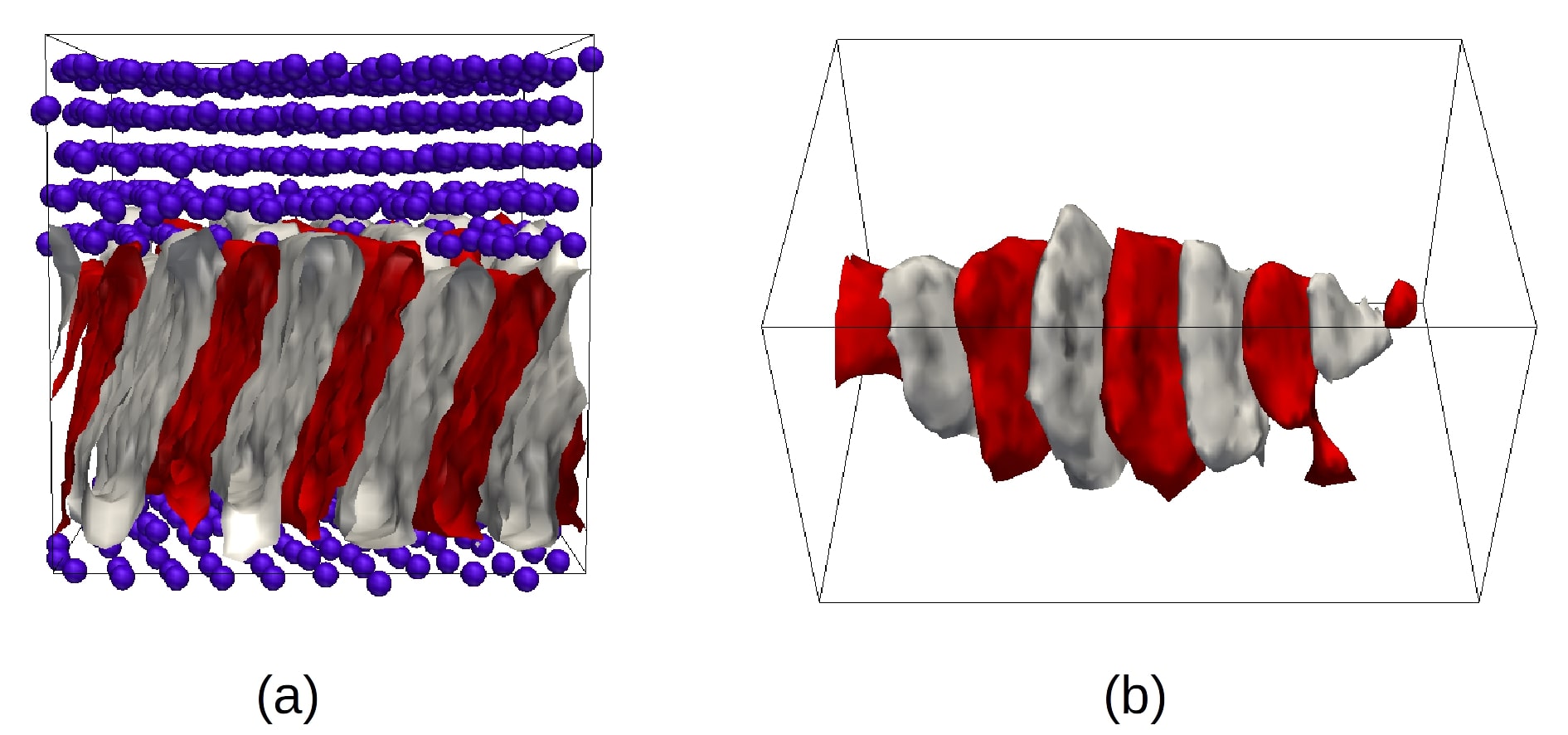}
\caption{
In (a) a simulation in a small box $V=32^3$ with a concentration $\phi_p=0.27$ and the same parameters as in Fig. \ref{fig:3d.lam-neutral}. 
In (b) we show a detail of Fig. \ref{fig:3d.lam-neutral} (c) without showing nanoparticles. 
}
\label{fig:3d.lam-neutral.details}
\end{figure}

Simulations shown in Figure \ref{fig:3d.lam-neutral} are the final snapshot for time $t/\tau_{pol} = 1.9 \times 10^5$ scaled with the diffusive time scale of the block copolymer. 
These are not the equilibrium configuration of the system, but a highly stable state of the evolution. 
For instance, NP clusters in Figure \ref{fig:3d.lam-neutral}  would eventually form a single NP cluster. 
Nonetheless, the segregation of NP clusters into BCP defects (lamellar grain boundaries) slows down the already slow pathway towards equilibrium.

\begin{figure}[hbtp]
\centering
\includegraphics[width=0.99\linewidth]{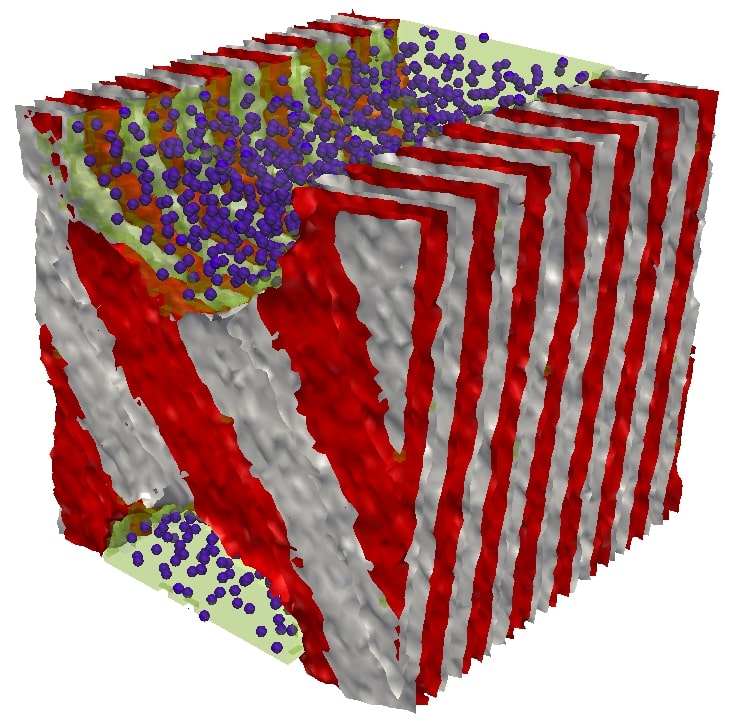}
\caption{
Simulation of a low concentration $\phi_p=0.1$ of neutral NPs sized $R_0=1.7$ in a symmetric $B=0.02$ BCP. 
NP-rich area -given by $-0.2<\psi < 0.2 $- is coloured transparent yellow.  
}
\label{fig:3d.lam-neutral.aggregation}
\end{figure}

In order to gain insight over the  equilibrium configuration of these co-assembled structures, we can perform a long-time calculation up to $t/\tau_{pol} =2.65 \times 10^6$  to obtain the morphology shown in Fig. \ref{fig:3d.lam-neutral.aggregation} (a). 
Here, we can observe a neutral NP-rich area -coloured in transparent yellow- where all NPs are phase-separated from the microphase-separated BCP lamellar domains.
In particular, due to periodic boundary conditions, we can identify a single NP-rich domain which has an elongated shape, in the direction of the lamellar interface (normal to the lamellar planes). 
Electronic supplementary material in Movie 1 shows the slow dynamics that results into this final state, which justifies the need for a highly efficient parallel computer code. 

\begin{figure}[hbtp]
\centering
\includegraphics[width=0.99\linewidth]{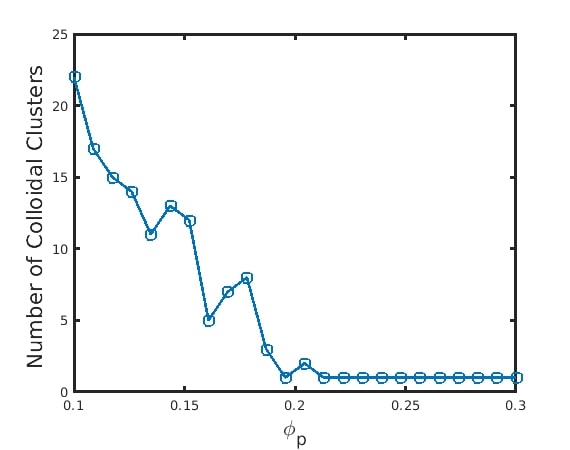}
\caption{Number of colloidal clusters formed by neutral colloids in a cylinder forming block copolymer mixture as a function of the concentration of particles.
 }
\label{fig:3d.cyl.iface}
\end{figure}

\begin{figure}[hbtp]
\centering
\includegraphics[width=0.99\linewidth]{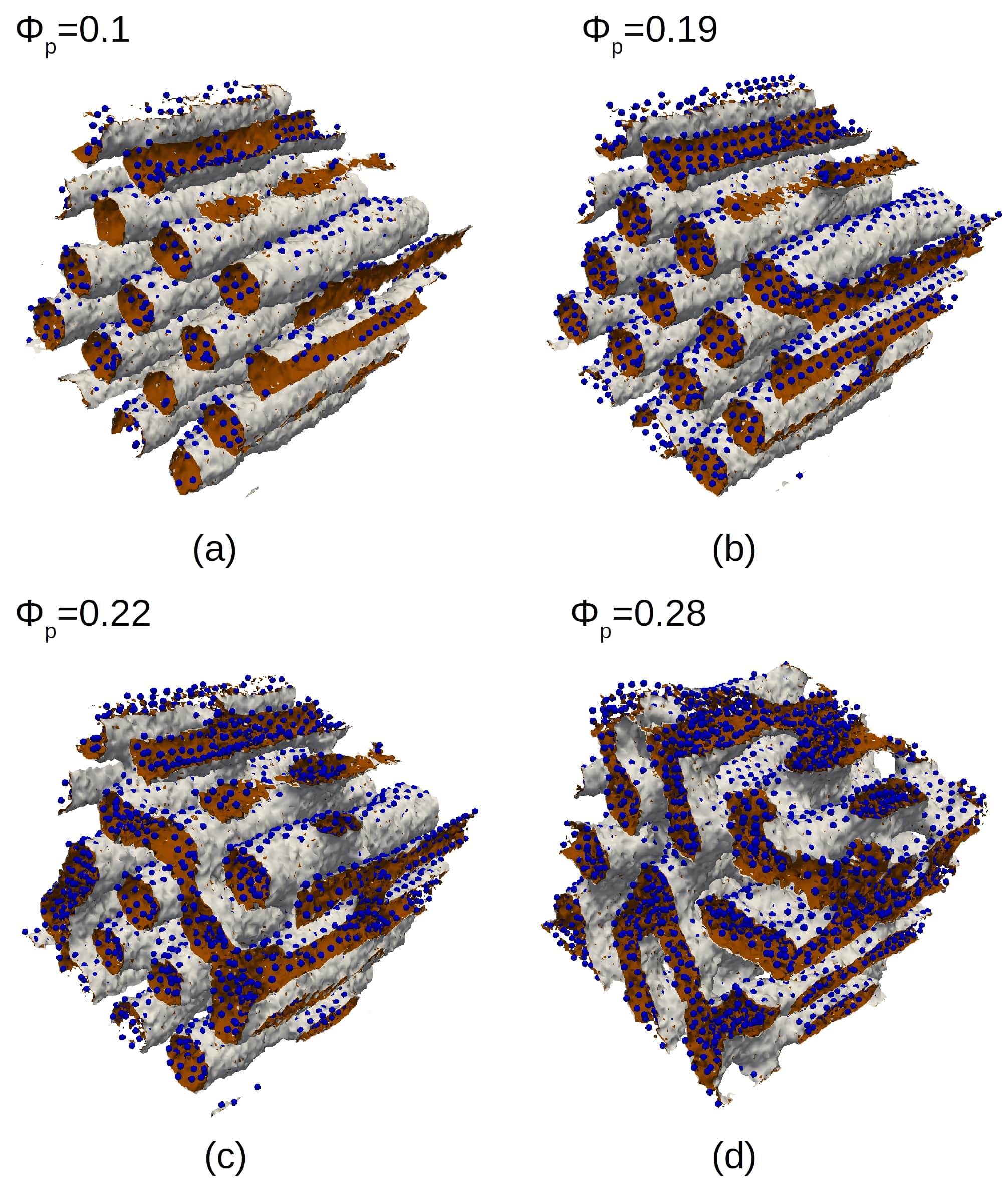}
\caption{
Snapshots of simulation results of neutral nanoparticles in a cylinder-forming block copolymer matrix corresponding to the curve in Figure \ref{fig:3d.cyl.iface}. Nanoparticles are segregated to the interface between red and grey domains. 
}
\label{fig:3d.cyl.iface.snaps}
\end{figure}

While the assembly of neutral nanoparticles at lamellar-forming has been experimentally studied , cylinder-forming (asymmetric) block copolymers/neutral nanoparticles have not been devoted experimental work.
In Figure \ref{fig:3d.cyl.iface}  we can observe the number of colloidal clusters in an asymmetrical ($f_0=0.35$), cylinder-forming BCP. 
At low concentrations colloids are found to simply be segregated within the interface of the cylinders. 
As the interface becomes saturated with  nanoparticles, the colloids start to form bridges along neighbouring domains. Figure \ref{fig:3d.cyl.iface.snaps} (a) we can observe the segregation of nanoparticles at interfaces. 
If the concentration of particles is higher than $\phi_p^* \sim 0.213$, nanoparticles form a single percolating cluster. 
Visual inspection in Figure \ref{fig:3d.cyl.iface.snaps} (b), (c) and (d) can draw the conclusion that the block copolymer maintains a phase-separated microstructure even at high concentrations as in $\phi_p=0.28$. 
This NP-induced morphology in the BCP can be related to that of a 2D asymmetric BCP mixed with neutral NPs, as in Figure 7 (a) in ref \cite{diaz_phase_2018}, where the majority-phase of the BCP forms a continuous, percolating lamellar-like domain, while the minority phase is enclosed by an NP-rich area which together with the minority BCP forms a lamellar-like percolating domain. 
Similarly, this bicontinuous BCP structure can be related to a lamellar-to-bicontinuous transition found in experiments \cite{kim_nanoparticle_2007}.

\subsection{Large nanoparticles}

\begin{figure*}[hbtp]
\centering
\includegraphics[width=0.9\textwidth]{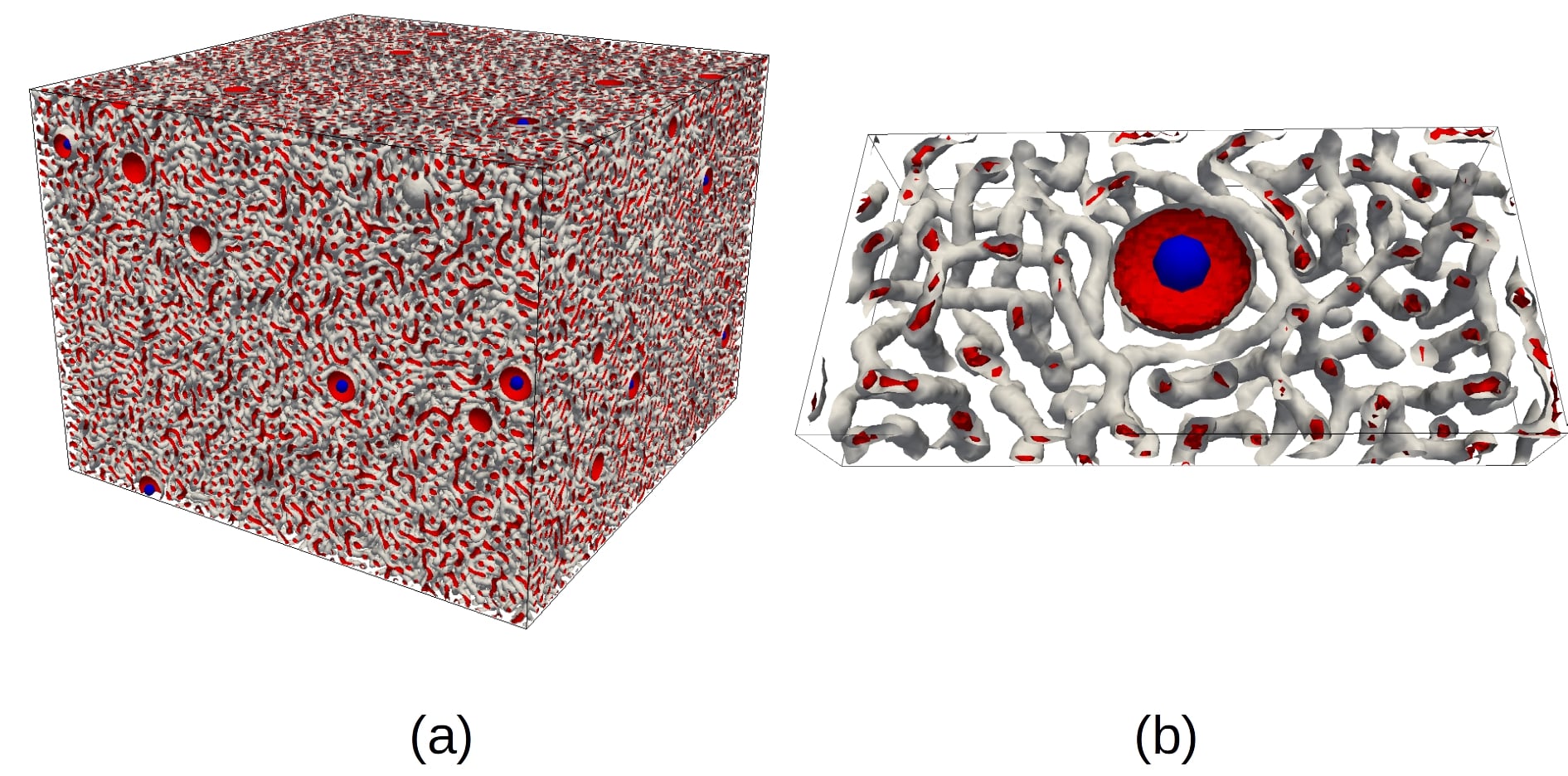}
\caption{Co-assembly of large NPs $R_0=8.5$ compatible with the minority block in a asymmetric $f_0=0.4$ BCP. System size is $V=400^2\times 300$ and $48$ processors were used.  }
\label{fig:3d.largeNP}
\end{figure*}

As an example on the ability of the presented model to scale up to considerably large systems, we can explore the regime in which the NP size is considerably larger than the BCP period. 
To this end we select parameters $R_0=8.5$ and $B=0.02$ for $N_p=100$ nanoparticles which have an affinity $\psi_0=-1$ in a cylinder-forming BCP $f_0=0.4$. 
The system size is $V=400^2\times 300$ with $n_p=4\times 4 \times 3$ processors. 
In Fig. \ref{fig:3d.largeNP} (a) the cylinder-forming morphology of the BCP can be observed, along with NPs which create a local perturbation in the nearby BCP. 
A detail of the BCP assembly near a large NP can be observed in (b), where we can find a spherical shell of the compatible block to the NP, followed by a secondary shell made of the minority copolymer forming cylinders. 
This behaviour can be directly related to BCP nanoshells, as in Fig. 2 (f) in Reference \cite{pinna_block_2008}. 

\section{Conclusions}

Three dimensional simulations of block copolymer/nanoparticle mixtures have been used to analyse the morphological changes induced by  colloids in a block copolymer melt, as well as the assembly of colloids within the phase-separated block copolymer. 
The comparatively fast parallel computational model has allowed us to achieve a vast range of number of particles, in turn making it possible to study the high concentration regime of particles. 

The order-to-order phase transition due to the presence of A-compatible nanoparticles in a symmetric, lamellar-forming BCP has been studied in detail in terms of the particle loading. 
It has been shown that the lamellar-to-cylinder transition evolves through a bicontinuous intermediate state that concludes in the well-defined cylindrical morphology. 
The time evolution towards this final structure can be tracked with the Euler characteristic, finding intermediate stages of connected networks of elongated domains. 

The assembly of colloids in the block copolymer has been studied both in the case of block-A compatible  and interface-compatible colloids. 
In the first case, we have studied the soft-confinement case in which the NP's ability to diffuse is strongly influenced by the lamellar phase separation. 
In fact, a layered, hexagonally close-packed assembly of colloids can be found which is driven by the relative length scales between the nanoparticle size and the lamellar spacing. 
Colloids are found to organize in crystal-like structures forming layers due to the confinement exerted by the BCP. 
This behaviour is non-monotonic with the number of layers dictated by the NP concentration and intermediate disordered colloidal states between defined layers.

Neutral, interface-compatible nanoparticles have been found to segregate to the surface between A-\textit{b}-B domains. A high concentration of nanoparticles at the interface tends to form NP-rich areas, which in the case of a symmetric lamellar-forming BCP results in macrophase separation of the BCP and NPs. 
At high concentration NPs tend to form bridges along  cylindrical domains, eventually forming a continuous network of nanoparticles. 
These complex morphologies are due to the presence of colloids and cannot be related to pure BCP phases (differently from lamellar to cylinders phase transitions). 
Comparisons can be drawn between the morphologies described in Fig. \ref{fig:3d.lam-neutral} and computer simulations of ternary blends made of BCP/homopolymer as described by Ref. \cite{ohta_dynamics_1995}. 

In this work we have made use of a highly efficient parallel code that can achieve system sizes and time scales which were previously unavailable. 
For instance, previous CDS/Brownian Dynamics three dimensional simulations\cite{ginzburg_three-dimensional_2002} reported simulation box of $64^3$ while Figure \ref{fig:3d.largeNP} uses $400^2\times 300$. 
Coarse grained methods such as DPD have reported up to a few BCP periods, while Figure \ref{fig:3d.largeNP} is able to simulate a system size of $50^2\times 37$ BCP periods. 
Additionally, this CDS scheme can reach considerably long time scales as shown in the formation of a single NP aggregate in Figure \ref{fig:3d.lam-neutral.aggregation}.

\section{Conflicts of interest}
There are no conflicts to declare.

\section{Acknowledgements}

The work has been performed under the Project HPC-EUROPA3 (INFRAIA-2016-1-730897), with the support of the EC Research Innovation Action under the H2020 Programme; in particular, the authors gratefully acknowledges the computer resources and technical support provided by Barcelona Supercomputing Center (BSC).
I. P. acknowledges support from MINECO (Grant No. PGC2018-098373-B-100), DURSI (Grant No. 2017 SGR 884) and SNF Project No. 200021-175719.

\bibliography{references}
\bibliographystyle{rsc} 

\end{document}